# COVID-19 DIAGNOSIS FROM X-RAY USING NEURAL NETWORKS

Dinesh J, Mohammed Rhithick A

Department of Software Engineering,

SRM Institute of Science and Technology, Chennai

**ABSTRACT:**

Corona virus or COVID-19 is a pandemic illness, which has influenced more than million of causalities worldwide and infected a few large number of individuals .Innovative instrument empowering quick screening of the COVID-19 contamination with high precision can be critically useful to the medical care experts. The primary clinical device presently being used for the analysis of COVID-19 is the Reverse record polymerase chain response as known as RT-PCR, which is costly, less-delicate and requires specific clinical work force. X-Ray imaging is an effectively available apparatus that can be a great option in the COVID-19 conclusion. This exploration was taken to examine the utility of computerized reasoning in the quick and exact recognition of COVID-19 from chest X-Ray pictures. The point of this paper is to propose a procedure for programmed recognition of COVID-19 from advanced chest X-Ray images applying pre-prepared profound learning calculations while boosting the discovery exactness. The point is to give over-focused on clinical experts a second pair of eyes through a learning picture characterization models. We distinguish an appropriate Convolutional Neural Network-CNN model through beginning similar investigation of a few mainstream CNN models. Such a tool can gauge severity of COVID-19 lung infections (and pneumonia in general) that can be used for escalation or de-escalation of care as well as monitoring treatment efficacy, especially in the ICU.

# INTRODUCTION:

Corona virus which began with the unknown cases of pneumonia in China on December 31,2019. This has now become a pandemic disease across the world. It has caused an overwhelming impact on both everyday lives, public health, and also world's economy. It is good to detect the positive cases on right time and could be even expected in order to prevent rapid spread to treat influenced patients. The need for the diagnostic tools has increased as there are no toolkits available. Severe of this disease affect respiratory system which leads to death. The most common symptoms of COVID 19 includes fever, cough, headache, dry throat, body pain, breath shortness.

According to the current situation most of the countries are undergoing 2nd wave of COVID 19. From the reports of this wave most of the researchers say that this disease may affect people above 18 to below 45. In future 3rd wave of COVID 19 can also be a threat for the peoples across the country which may also affect children's above age 12 to 45.

Totally, according to WHO results as of 26 may 2021 there are 167,492,769 confirmed cases of COVID-19 which includes 3,482,907 deaths.

**Related work:**

Coronavirus disease 2019 also known as COVID-19 has become a pandemic. The disease is caused by a beta coronavirus called Severe Acute Respiratory Syndrome If the diagnosis is fast-paced, the disease can be controlled in a better manner. Laboratory tests are available for diagnosis, but they are bounded by available testing kits and time. Specifically, chest X-Ray images can be analysed to identify the presence of COVID-19 in a patient. -In this paper, an automated method for the diagnosis of COVID-19 from the chest X-Ray images is proposed.We are using improved CNN model for analysing the chest X-RAY.All the inputs are fed into a network for analysing the disease.The input In the network has been designed to predict whether XRAY states Covid or normal.This method is best suitable for the current condition which can be used for the effective diagnosis of the disease.

Pre-trained VGG-16 with CNN Architecture to classify X-Rays images into Normal or Pneumonia. Chest radiograph (x-ray) are primarily used to diagnose pneumonia However, x-ray examination is a difficult operation, even for a qualified radiologist and there is a need to improve the accuracy of the diagnosis.In this paper we study that VGG16 is one the most finest algorithm which shows accurate accuracy and this model in CNN has showed greater results in validation and prediction. So using VGG16 we are predicting pneumonia or normal of chest XRay image. To refine the training samples in a balanced way we have used Data augmentation. This method is best technique to find out best accuracy of the datatset.This Model has successfully implemented in python using Keres library.]Our Pt-Net is a novel object detection network based on a pre-trained and multi-feature VGG-16 network. Firstly, Pt-Net is initialized by a pre-trained VGG-16 model and its own CNN output via a linear combination. Secondly we use Pt-Net which generates particle filter method on Conv5 which helps us to combine two images and will be used for best prediction. After that we use multi layer process for cropped images and insert a two dimensional overlap function loss for localization. At last we apply Pt-Net for both the face detection and object detection which is well trained with Wider face dataset and Pascal voc dataset. Finally Pt-Net produces best results on prediction of object detection.COVID-19 Detection in X-ray Images using CNN Algorithm. In this paper we developed an algorithm which is known as CheXNet algorithm. This CheXNet was developed to detect and diagnose pneumonia using Chest Xray dataset.to achieve some better performance radiologists have made simple changes to the algorithm which helps to diagnose 14 pathological conditions in the Xray which can show better performance than previously developed Deep learning models. By this algorithm we have experimented by applying Convolutional neural network in a similar way to the CheXNet algorithm by providing 500 datasets of chest Xray images collected from Kaggle websites. Finally it stated that some of the datasets are affected with COVID and some of them are normal. Prediction accuracy states that this accuracy is close to the results of CheXNet algorithm. Can AI Help in Screening Viral and COVID-19 Pneumonia- Coronavirus disease (COVID-19) is a pandemic disease, which has already caused thousands of causalities and infected several millions of people worldwide. The main clinical tool currently in use for the diagnosis of COVID-19 is the Reverse transcription polymerase chain reaction. In this paper we have used Xray imaging which can easily accessible tool that can be an excellent alternative in the COVID-19 diagnosis. The main aim is to provide a better technique from Chest Xray images which is trained using Deep learning method which can predict better accuracy. All the images are stored in the network as input. Where network is classifies into two prection results Covid or Normal. Finally after predicting the results It shows good accuracy and speed of the accuracy value is increased which will be very useful for this pandemic condition.

## Proposed work:

- We develop an algorithm that can detect pneumonia and Covid-19 from chest X-rays at a level of practicing radiologists.
- We use the chest x-ray images that are available for the open source from covid-dataset and combine them into dataset.
- We created out train and test data sets that consists of more than 3000 images. The images in the train and test data sets are properly labelled with "COVID-19" or "NORMAL".
- After training the model we will save as h5 file and use it for prediction of covid later in the program.

## System architectute:

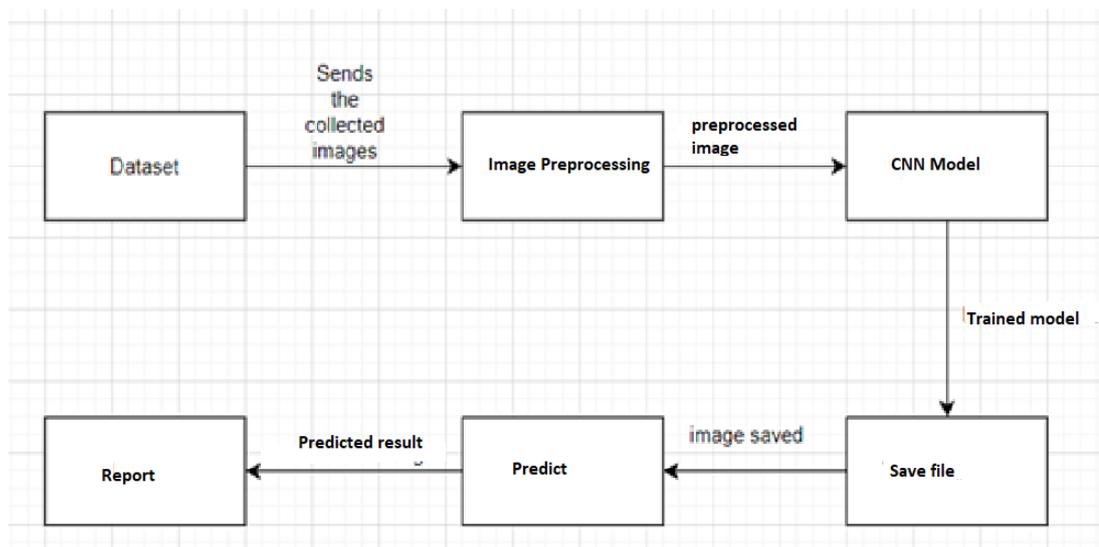

Fig 1

The system architecture defines the structure and complex framework of the system. It consists of the components which are combined gives the complete system. The given diagram picturizes the working of the system.

The following primary findings are frequently observed in the chest X-rays of COVID-19 patients .

- Ground-glass opacities (GGO) (bilateral, multifocal, subpleural, peripheral, posterior, medial and basal).

- A crazy paving appearance (GGOs and inter-/intra-lobular septal thickening).

- Air space consolidation.

- Bronchovascular thickening (in the lesion).

- Traction bronchiectasis.

## Experimental setup:

Here, first importing all the libraries which will be needed to implement our model. We will be using Sequential method as it's a sequence model.

Images from a public COVID-19 database were scored retrospectively by three blinded experts in terms of the extent of lung involvement as well as the degree of opacity. A neural network model that was pre-trained on large (non-COVID-19) chest X-ray datasets is used to construct features for COVID-19 images which are predictive for our task.

ImageDataGenerator is used from keras as it helps with many functions like rescale, rotate, zoom, flip, etc., Relu (Rectified Linear Unit) activated to each layers so that all the negative values are not passed to the next layer.

Here we will visualise plots of training/validation accuracy and loss using matplotlib and seaborn. After training the model we will save as h5 file and use it for prediction of covid later in the program.

## RESULT AND ANALYSIS:

First, we performed experiments for detecting COVID using X-ray images on different algorithm and models with different scenarios and weights. The main reason in experimenting with different algorithms and models is to see which one suits it better with good response time. However, when the keras model algorithm examined the images there were ups and downs in each epoch training season, which were resolved later.

Algorithms we used in our model like VGG16, RESNET, INCEPTION V3 for studying which algorithms fits best. But after the study with the data we concluded that our model works superior that the algorithm.

| S.NO | Algorithm/Model | Accuracy |
|---|---|---|
| 1 | VGG16 | 92.33 |
| 2 | RESNET | 90.33 |
| 3 | INCEPTION V3 | 89.44 |
| 4 | Convolutional layer with RMSprop optimizer | 97.99 |

The CNN based learning model contains 167,105 parameters. Weights are updated with the RMSprop optimizer. 70% of the images are used in training set and remaining are used for testing set.

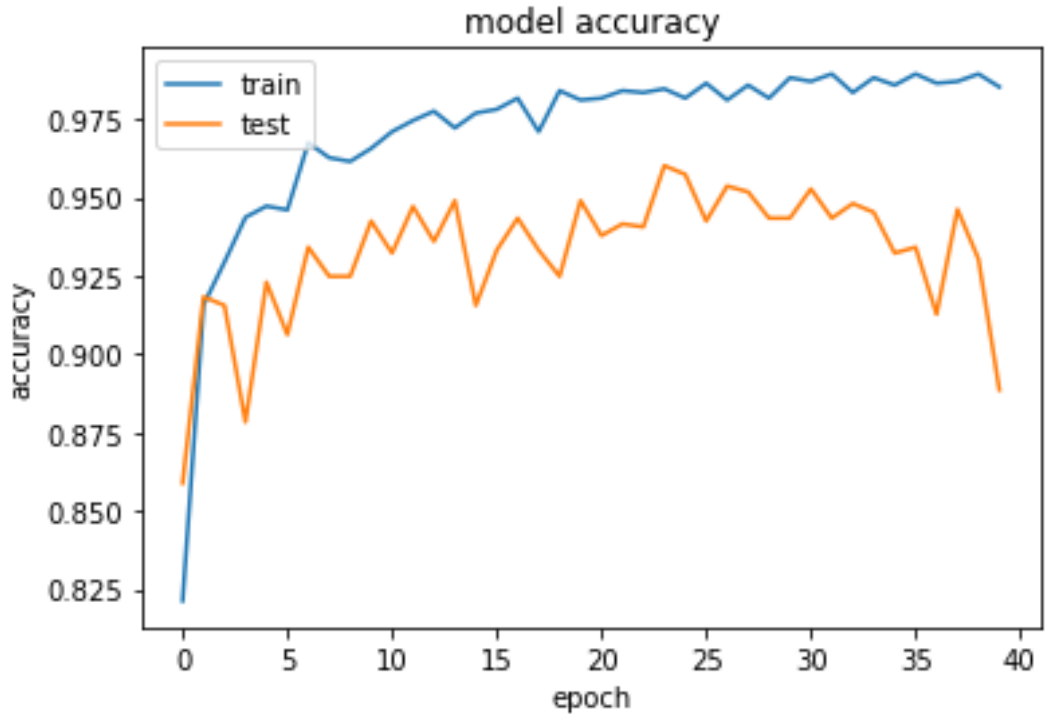

**Graph 1**

The model graph shows the plot with good accuracy of 98% .

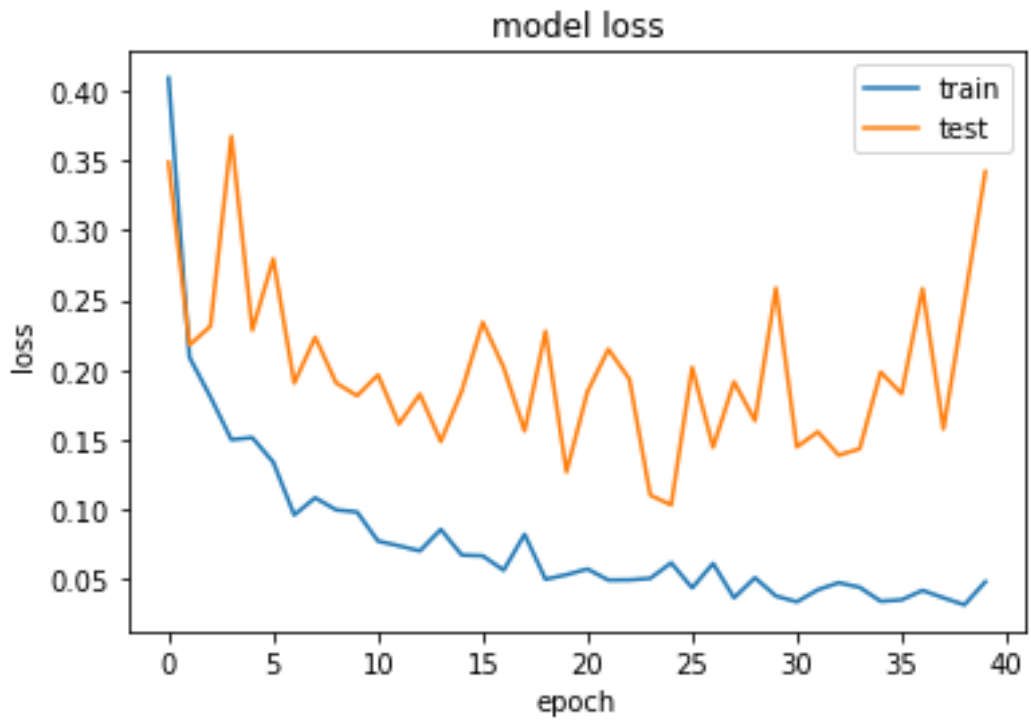

**Graph 2**

The objective of project is developing an application which helps in detection of COVID with faster and cheaper process than that are available in the market.

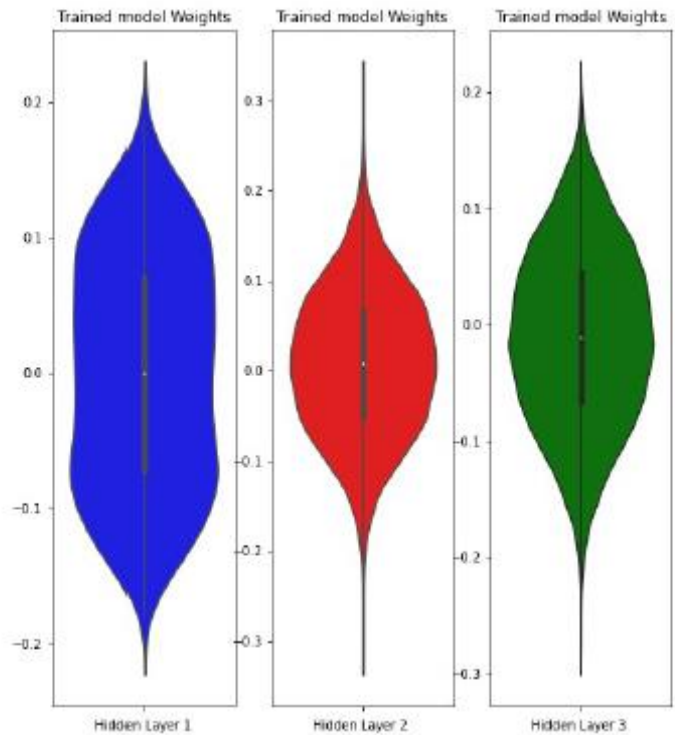

**Graph 3**

The most important constraint in the model is its accuracy. So we divided the class into two for binary classification which worked very well in the convolutional neural model.

**CONCLUSION AND FUTURE WORK**:

For today's small and difficult COVID-19 datasets, we have shown that the CNN model can be used to develop suitable deep learning-based tools for COVID-19 detection. This work presents a CNN-based deep transmission learning approach for the automatic detection of respiratory disease COVID-19.

In addition, due to the differences in experts' experience, the input images of x-ray equipment vary widely. Because its network is efficiently trained, artificial intelligence shows excellent performance in classifying COVID-19 respiratory diseases. From a huge data set. We tend to believe that this computer-assisted diagnostic tool will dramatically improve the speed and accuracy of detecting positive cases of COVID-19.

This tactic is useful during the pandemic, as long as the region is willing to take measures to prevent the disease burden and the resources are available.

9. A Comparison Based Breast Cancer High Microscopy Image Classification Using Pre-trained Models-(2020)-Nidhi Ojha; Ashwani Kumar

10. Pre-trained VGG-16 with CNN Architecture to classify X-Rays images into Normal or Pneumonia-(2019)-P Naveen; B Diwan